\documentclass[reprint,amsmath,amssymb]{revtex4-2}
\usepackage[utf8]{inputenc}
\usepackage[T1]{fontenc}
\usepackage{graphicx}
\usepackage[svgnames]{xcolor}
\usepackage{dcolumn}
\usepackage{bm}
\usepackage{enumerate}
\usepackage{hyperref}
\usepackage{mathcomp}
\hypersetup{
 colorlinks=true,%
 linkcolor=MidnightBlue,
 citecolor=MidnightBlue,
 urlcolor=MidnightBlue,
}
\everymath{\displaystyle}
\begin{document}

\preprint{}

\title{Dark energy in conformal Killing gravity}

\author{Junpei Harada}
 \email{jharada@hoku-iryo-u.ac.jp}
\affiliation{Health Sciences University of Hokkaido, 1757 Kanazawa, Tobetsu-cho, Ishikari-gun, Hokkaido 061-0293, Japan}

\date{November 8, 2023}

\begin{abstract}
The Friedmann equation, augmented with an additional term that effectively takes on the role of dark energy, is demonstrated to be an exact solution to the recently proposed gravitational theory named “conformal Killing gravity.” This theory does not explicitly incorporate dark energy. This finding suggests that there is no necessity to postulate the existence of dark energy as an independent physical entity. The dark energy derived from this theory is characterized by a specific equation of state parameter, denoted as $\omega$, which is uniquely determined to be $-5/3$. If this effective dark energy is present, typically around 5\% of the total energy density at the present time, and under the assumption of density parameters for matter and the cosmological constant, $\Omega_{\rm m}\sim 0.25$ and $\Omega_\Lambda \sim 0.7$, respectively, the expansion of the universe at low redshifts ($z < 1.5$) can exceed expectations, while the expansion at $z > 1.5$ remains unchanged. This offers a potential solution to the Hubble tension problem. Alternatively, effective dark energy could be a dominant component in the present-day universe. In this scenario, there is also the potential to address the Hubble tension, and furthermore, it resolves the coincidence problem associated with the cosmological constant.
\end{abstract}

\maketitle

\section{Introduction}
Recently, in the gravitational theory proposed by~\cite{Harada:2023rqw}, which is referred to as “conformal Killing gravity”~\cite{Mantica:2023}, the Friedmann equation has been generalized as follows:
\begin{equation}
	2\left(\frac{\dot{a}(t)}{a(t)}\right)^2 - \frac{\ddot{a}(t)}{a(t)} = \frac{4\pi G}{3}(5\rho(t) + 3p(t)) - \frac{2k}{a^2(t)} + \frac{\Lambda}{3}.
	\label{eq:Harada_Friedmann}
\end{equation}
Here, $a(t)$ represents the scale factor, the dots denote the time derivative, $\rho(t)$ and $p(t)$ stand for the energy density and pressure, respectively, and $k$  is a constant representing the curvature of three-dimensional space. The cosmological constant $\Lambda$ in Eq.~\eqref{eq:Harada_Friedmann} is derived as an integration constant~\cite{Harada:2023rqw,Mantica:2023}. Equation (1) has been independently derived through two distinct methods~\cite{Harada:2023rqw,Mantica:2023}. Very recently, Barnes~\cite{Barnes:2023uru} discovered the most general static spherically symmetric solution in that gravitational theory.

The Eq.~\eqref{eq:Harada_Friedmann} demonstrates an intriguing property~\cite{Harada:2023rqw,Mantica:2023}: Despite the absence of negative pressure or the cosmological constant, the universe described by Eq.~\eqref{eq:Harada_Friedmann}  undergoes a transition from decelerating to accelerating expansion. To illustrate this cosmological transition, a solution for the scale factor $a(t)$ was derived within a matter-dominated universe~\cite{Harada:2023rqw}. The same solution was obtained through a different approach~\cite{Mantica:2023}. This solution explicitly describes the transition from deceleration to acceleration. Remarkably, this was achieved without the need for negative pressure or a positive cosmological constant $\Lambda$.

In contrast to the previous study~\cite{Harada:2023rqw}, which focused solely on a matter-dominated universe, this work removes such constraints. Instead, we consider various components of the universe, including matter (m), radiation (r), curvature ($k$), and the cosmological constant ($\Lambda$). Throughout this study, we do not explicitly assume the existence of dark energy as a distinct physical entity. We demonstrate that Eq.~\eqref{eq:Harada_Friedmann}, even without the inclusion of any dark energy, is equivalent to the standard Friedmann equation that incorporates a specific form of dark energy. In our theory, dark energy is merely an effective concept, and it does not represent a distinct physical entity. This perspective differs from that of general relativity, where dark energy is typically regarded as a real physical entity.

The dark energy derived from this approach possesses an equation of state parameter, $\omega=p/\rho$, which is uniquely determined to be $-5/3$. Just a few days ago, Mantica and Molinari reported the same result~\cite{Mantica:2023}. If this effective dark energy constitutes approximately 5\% of the total energy density at the present time, and given the density parameters for matter and the cosmological constant, $\Omega_{\rm m}\sim 0.25$ and $\Omega_\Lambda \sim 0.7$ respectively, the expansion of the universe at low redshifts ($z < 1.5$) can exceed expectations, while the expansion at $z > 1.5$ remains unaffected. As a result, this has a potential to address the Hubble tension problem~\cite{Riess:2019qba,DiValentino:2021izs,Dainotti:2021pqg,Dainotti:2022bzg,Kamionkowski:2022pkx,Lenart:2022nip,Bargiacchi:2023jse,Dainotti:2023ebr,Bargiacchi:2023rfd}.

There is an even more intriguing possibility to consider: the current universe could be matter-dominated with a vanishing cosmological constant. We show that even in a matter-dominated universe with $\Lambda=0$, the Hubble tension can potentially be resolved. In this scenario, the coincidence problem associated with the cosmological constant is eliminated because the cosmological constant is zero. This is theoretically an advantage compared to models with a nonzero $\Lambda$.

This paper is organized as follows: In Sec.~\ref{sec:effective}, we demonstrate that Eq.~\eqref{eq:Harada_Friedmann}, even when not incorporating any dark energy, is equivalent to the Friedmann equation that includes a specific type of dark energy. Section~\ref{sec:omega} presents another derivation for $\omega=-5/3$. In Sec.~\ref{sec:Hubble}, we investigate the potential of effective dark energy to resolve the Hubble tension problem. Finally, in Sec.~\ref{sec:conclusion}, we provide a summary and conclusions.

 \section{effective dark energy\label{sec:effective}}
 We assume that the universe is composed of matter (m) and radiation (r), and we do not postulate the existence of dark energy. Therefore, the energy density $\rho$ and pressure $p$ in Eq.~\eqref{eq:Harada_Friedmann} are given by $\rho=\rho_{\rm m} + \rho_{\rm r}$ and $p=\rho_{\rm r}/3$, respectively. In this case, Eq.~\eqref{eq:Harada_Friedmann} takes the form of 
 \begin{equation}
	2\left(\frac{\dot{a}}{a}\right)^2 - \frac{\ddot{a}}{a} = \frac{4\pi G}{3}(5\rho_{\rm m} + 6\rho_{\rm r}) - \frac{2k}{a^2} + \frac{\Lambda}{3}.
	\label{eq:Harada_Friedmann3}
\end{equation}
Here, the energy density $\rho_{\rm m}$ and $\rho_{\rm r}$ as functions of $a$ can be derived from the conservation law $\nabla_\mu T^\mu{}_{\nu}=0$, 
\begin{eqnarray}
	0=-\nabla_\mu T^\mu{}_0 = \dot{\rho}+3\frac{\dot{a}}{a}(\rho + p).
	\label{eq:conservation}
\end{eqnarray}
Assuming $p=\omega \rho$ with $\omega$ time-independent, Eq.~\eqref{eq:conservation} gives $\rho \propto a^{-3(1+\omega)}$. This provides expressions for matter ($\omega=0$) and radiation $(\omega=1/3)$ as follows:
\begin{subequations}
\begin{eqnarray}
	\rho_{\rm m} (t) &= \rho_{\rm m,0}\left(\frac{a(t)}{a_0}\right)^{-3},\\
	\rho_{\rm r} (t) &= \rho_{\rm r,0}\left(\frac{a(t)}{a_0}\right)^{-4},
\end{eqnarray}
\end{subequations}
where $\rho_{\rm m,0}$ and $\rho_{\rm r,0}$ represent the density for matter (m) and radiation (r) at the present time, respectively. The $a_0$ denotes the scale factor at the present time.

Using the Hubble parameter and its time derivative,
\begin{eqnarray}
	H \equiv \frac{\dot{a}}{a},\quad
	\dot{H}=\frac{\ddot{a}}{a} - H^2,
\end{eqnarray}
the left-hand side of Eq.~\eqref{eq:Harada_Friedmann3} can be expressed as 
\begin{eqnarray}
	2\left(\frac{\dot{a}}{a}\right)^2 - \frac{\ddot{a}}{a} = H^2-\dot{H}.
	\label{eq:lhs_GFE3}
\end{eqnarray}
Substituting Eq.~\eqref{eq:lhs_GFE3} into Eq.~\eqref{eq:Harada_Friedmann3} and dividing by $H_0^2$, we find that Eq.~\eqref{eq:Harada_Friedmann3} can be expressed as 
\begin{eqnarray}
	&& \frac{H^2 - \dot{H}}{H_0^2}  \nonumber\\
	=
	&& \frac{5}{2} \Omega_{\rm m} \left(\frac{a}{a_0}\right)^{-3}
	+3 \Omega_{\rm r} \left(\frac{a}{a_0}\right)^{-4}
	+2 \Omega_k \left(\frac{a}{a_0}\right)^{-2} 		
	+\Omega_\Lambda.\nonumber\\
	\label{eq:Harada_Friedmann4}
\end{eqnarray}
Here, the density parameters $\Omega$'s are defined as follows: 
\begin{eqnarray}
	\Omega_{\rm m} \equiv \frac{\rho_{\rm m,0}}{\rho_{\rm c}},\
	\Omega_{\rm r}  \equiv \frac{\rho_{\rm r,0}}{\rho_{\rm c}},\
	\Omega_k \equiv - \frac{k}{a_0^2 H_0^2},\
	\Omega_\Lambda \equiv \frac{\Lambda}{3H_0^2},\qquad
\end{eqnarray}
and the critical density is defined as $\rho_{\rm c} \equiv 3H_0^2/8\pi G$. Equation~\eqref{eq:Harada_Friedmann4} does not contain dark energy, since we do not assume the presence of dark energy. Thus, Eq.~\eqref{eq:Harada_Friedmann4} includes only four components: $\Omega_{\rm m}$, $\Omega_{\rm r}$, $\Omega_k$, and $\Omega_\Lambda$.

These four density parameters do not necessarily satisfy the relation, $\Omega_{\rm m} + \Omega_{\rm r} + \Omega_k + \Omega_\Lambda=1$, which represents the Friedmann equation at the present time. Instead, they satisfy the following relation:
\begin{equation}
	2+q_0=\frac{5}{2}\Omega_{\rm m} + 3\Omega_{\rm r} + 2 \Omega_k + \Omega_\Lambda.
	\label{eq:q0_Omega1}
\end{equation}
Here, the deceleration parameter $q$ is defined by
\begin{eqnarray}
	q\equiv - \frac{\ddot{a}a}{\dot{a}^2} = -\frac{\ddot{a}}{aH^2}=-\frac{\dot{H}}{H^2}-1,
	\label{def:deceleration}
\end{eqnarray}
and $q_0$ represents its present value. Equation~\eqref{eq:q0_Omega1} can be derived as follows. Using Eq.~\eqref{def:deceleration}, the left-hand side of Eq.~\eqref{eq:Harada_Friedmann4} can be expressed as
\begin{eqnarray}
	\frac{H^2-\dot{H}}{H_0^2} = (2+q)\left(\frac{H}{H_0}\right)^2.
	\label{eq:lhs_GFE4}
\end{eqnarray}
Substituting Eq.~\eqref{eq:lhs_GFE4} into Eq.~\eqref{eq:Harada_Friedmann4} and then taking the present value, we obtain Eq.~\eqref{eq:q0_Omega1}. 

In general relativity, four density parameters satisfy the relation $\Omega_{\rm m} + \Omega_{\rm r} + \Omega_k + \Omega_\Lambda=1$. In this case, Eq.~\eqref{eq:q0_Omega1} reads
\begin{eqnarray}
		q_0=\frac{1}{2}\Omega_{\rm m} + \Omega_{\rm r} - \Omega_\Lambda.
	\label{eq:q0_GR}
\end{eqnarray}
Equation~\eqref{eq:q0_GR} indicates that $q_0$ can take a negative value only when $\Omega_\Lambda > \Omega_{\rm m}/2+\Omega_{\rm r}$. When $\Omega_\Lambda=0$, $q_0$ is necessarily positive, which implies the decelerating expansion of the current universe.

In our theory, the result differs from general relativity. The sum of the four density parameters, $\Omega_{\rm m} + \Omega_{\rm r} + \Omega_k + \Omega_\Lambda$, does not necessarily equal one. Instead,  Eq.~\eqref{eq:q0_Omega1} should be satisfied. Equation~\eqref{eq:q0_Omega1} indicates that $q_0$ can have a negative value if the following relation is satisfied:
\begin{eqnarray}
	\frac{5}{2}\Omega_{\rm m} + 3\Omega_{\rm r} + 2 \Omega_k + \Omega_\Lambda < 2.
\end{eqnarray}
In particular, $q_0$ can become negative even when $\Omega_\Lambda=0$. For instance, in the case of a matter-dominated universe where $\Omega_{\rm r}=\Omega_k=\Omega_\Lambda=0$, $q_0$ is negative if $\Omega_{\rm m} <0.8$ (This outcome is consistent with a previous study~\cite{Harada:2023rqw}). Consequently, within the cosmological framework described by Eq.~\eqref{eq:Harada_Friedmann4}, the present-day expansion of the universe can be accelerating ($q_0<0$), all without the necessity of negative pressure or a cosmological constant.

An explicit solution for the scale factor $a(t)$ was obtained~\cite{Harada:2023rqw} by assuming a matter-dominated universe with $\Omega_{\rm r}=\Omega_k=\Omega_\Lambda=0$. Recently, the same solution was independently derived in another study~\cite{Mantica:2023}. This solution describes the transition from decelerated to accelerated expansion. In the subsequent discussion, we will clearly explain the mechanisms that facilitate this acceleration.

Equation~\eqref{eq:Harada_Friedmann4} includes a time derivative term, $\dot{H}$, makes it into a differential equation for the Hubble parameter. When we solve Eq.~\eqref{eq:Harada_Friedmann4} for $H$, we obtain
\begin{eqnarray}
	\left(\frac{H}{H_0}\right)^2 
	= &&\Omega_{\rm m}\left(\frac{a}{a_0}\right)^{-3} 
	+ \Omega_{\rm r}\left(\frac{a}{a_0}\right)^{-4}
	+ \Omega_k\left(\frac{a}{a_0}\right)^{-2}  \nonumber\\
	&&+ \Omega_\Lambda
	+ (1-\Omega_{\rm m}-\Omega_{\rm r}-\Omega_k-\Omega_\Lambda)\left(\frac{a}{a_0}\right)^2.\qquad
	\label{eq:Hubble}
\end{eqnarray}
For convenience, we define the coefficient in Eq.~\eqref{eq:Hubble} as
\begin{eqnarray}
	\Omega_{\rm eff} \equiv 1 - \Omega_{\rm m} - \Omega_{\rm r} - \Omega_k - \Omega_\Lambda.
	\label{eq:five_parameters}
\end{eqnarray}
Here, the “eff” subscript stands for “effective,” and its significance will become clear shortly. 

Remarkably, Eq.~\eqref{eq:Hubble} is an exact solution to Eq.~\eqref{eq:Harada_Friedmann4}. This can be readily confirmed as follows: By differentiating Eq.~\eqref{eq:Hubble} with respect to $t$ and subsequently dividing the result by $-2H$, we obtain
\begin{eqnarray}
	-\frac{\dot{H}}{H_0^2} 
	=&& \frac{3}{2}\Omega_{\rm m}\left(\frac{a}{a_0}\right)^{-3} 
		+ 2 \Omega_{\rm r}\left(\frac{a}{a_0}\right)^{-4}\nonumber\\
		&&+ \Omega_k\left(\frac{a}{a_0}\right)^{-2}
	-\Omega_{\rm eff}\left(\frac{a}{a_0}\right)^2.
	\label{eq:Hubbledot}
\end{eqnarray}
We find that the sum of Eqs.~\eqref{eq:Hubble} and~\eqref{eq:Hubbledot} is equal to Eq.~\eqref{eq:Harada_Friedmann4}. Thus, Eq.~\eqref{eq:Hubble} is an exact solution to Eq.~\eqref{eq:Harada_Friedmann4}.

Equation~\eqref{eq:Hubble} is a generalization of the Friedmann equation, augmented by an additional term, $\Omega_{\rm eff}(a/a_0)^2$. This term is characterized by the exponent 2, a value uniquely determined as a consequence of Eq.~\eqref{eq:Harada_Friedmann4}. Referring to Eq.~\eqref{eq:conservation}, we can deduce that $2=-3(1+\omega)$, which leads to the conclusion that $\omega$ equals $-5/3$. Consequently, the term $\Omega_{\rm eff}(a/a_0)^2$ effectively takes on the role of dark energy with $\omega=-5/3$. Its density parameter is given by $\Omega_{\rm eff}\equiv 1-\Omega_{\rm m}-\Omega_{\rm r}-\Omega_k-\Omega_\Lambda$.

Let us provide some clarifications here. First, the energy density $\rho$ is composed of matter (m) and radiation (r), as explicitly shown in Eq.~\eqref{eq:Harada_Friedmann3}. The extra term $\Omega_{\rm eff}(a/a_0)^2$ in Eq.~\eqref{eq:Hubble} effectively serves the role of dark energy, despite the absence of dark energy as an independent physical entity. Instead, it behaves analogously to dark energy. This perspective differs notably from the one in general relativity, where dark energy is conventionally regarded as an actual physical entity. Second, the parameter $\omega$ for the effective component is uniquely determined as $-5/3$. This value is not arbitrary. We will provide another straightforward derivation for $\omega=-5/3$ in Sec.~\ref{sec:omega}.

\section{Another derivation for $\omega=-5/3$\label{sec:omega}}

As pointed out in Ref.~\cite{Mantica:2023}, our gravitational field equation proposed in Ref.~\cite{Harada:2023rqw} can be expressed as
\begin{eqnarray}
	&&\nabla_\rho K_{\mu\nu} + \nabla_\mu K_{\nu\rho} + \nabla_\nu K_{\rho\mu}\nonumber\\
	&&-\frac{1}{6}(g_{\mu\nu}\partial_\rho + g_{\nu\rho}\partial_\mu + g_{\rho\mu}\partial_\nu)K = 0,
	\label{eq:GFE_Mantica}
\end{eqnarray}
where $K\equiv K^\mu{}_\mu$ and the tensor $K_{\mu\nu}$ is defined by
\begin{eqnarray}
	8\pi GK_{\mu\nu} \equiv R_{\mu\nu} - \frac{1}{2}Rg_{\mu\nu} + \Lambda g_{\mu\nu} - 8\pi G T_{\mu\nu}. 
	\label{eq:tensor_K}
\end{eqnarray}

Without loss of generality, we can absorb the cosmological term $\Lambda g_{\mu\nu}$ in Eq.~\eqref{eq:tensor_K} into the definition of $K_{\mu\nu}$. This is because the cosmological terms vanish in Eq.~\eqref{eq:GFE_Mantica}. This implies that $\Lambda$ in Eq.~\eqref{eq:tensor_K} is an integration constant. For the sake of convenience, we will employ the definition of Eq.~\eqref{eq:tensor_K}. Equation~\eqref{eq:GFE_Mantica} represents that the tensor $K_{\mu\nu}$ is a divergence-free gradient conformal Killing tensor~\cite{Mantica:2023}. Indeed, Eq.~\eqref{eq:GFE_Mantica} guarantees $\nabla_\mu K^\mu{}_\nu=0$. One can easily check it by contracting arbitrary two indices of Eq.~\eqref{eq:GFE_Mantica}. From Eq.~\eqref{eq:tensor_K}, using $\nabla_\mu K^\mu{}_\nu=0$ and the Bianchi identity, we obtain the conservation law $\nabla_\mu T^\mu{}_\nu=0$.

The tensor $K_{\mu\nu}$ defines the right-hand side of Eq.~\eqref{eq:tensor_K}. In the present case where the universe is assumed to be isotropic and spatially homogeneous, the components of the tensor $K_{\mu\nu}$ can be expressed as
\begin{eqnarray}
	K^\mu{}_\nu = {\rm diag}(-\rho_{\rm eff}(t), p_{\rm eff}(t), p_{\rm eff}(t), p_{\rm eff}(t))
\end{eqnarray}
and its trace is
\begin{eqnarray}
	K\equiv K^\mu{}_\mu = -\rho_{\rm eff}(t) + 3 p_{\rm eff}(t).
\end{eqnarray}
The conservation law $\nabla_\mu K^\mu{}_\nu=0$ gives
\begin{eqnarray}
	0 = -\nabla_\mu K^\mu{}_0 = \dot{\rho}_{\rm eff} + 3 \frac{\dot{a}}{a}(\rho_{\rm eff}+p_{\rm eff}).
	\label{eq:divergence_K}
\end{eqnarray}
General relativity corresponds to the special case where $\rho_{\rm eff}=p_{\rm eff}=0$.

Equation~\eqref{eq:tensor_K} is the same form of the Einstein equation with the substitution $T_{\mu\nu} \rightarrow T_{\mu\nu}+K_{\mu\nu}$. Consequently, Eq.~\eqref{eq:tensor_K} can be formulated as the Friedmann equations:
\begin{eqnarray}
	\left(\frac{\dot{a}}{a}\right)^2&=&\frac{8\pi G}{3}(\rho_{\rm m}+\rho_{\rm r}+\rho_{\rm eff})	
	-\frac{k}{a^2} + \frac{\Lambda}{3},\label{eq:Friedmann_a}
	\\
	\frac{\ddot{a}}{a}&=&-\frac{4\pi G}{3}\left(\rho_{\rm m} + 2\rho_{\rm r} + (1+3\omega)\rho_{\rm eff}\right)+ \frac{\Lambda}{3},
	\label{eq:Friedmann_b}
\end{eqnarray}
where we have used $p_{\rm m}=0$, $p_{\rm r}=\rho_{\rm r}/3$, and $p_{\rm eff} \equiv \omega \rho_{\rm eff}$. 

Substituting Eqs.~\eqref{eq:Friedmann_a} and~\eqref{eq:Friedmann_b} into Eq.~\eqref{eq:Harada_Friedmann3}, we obtain
\begin{eqnarray}
	(5+3\omega) \rho_{\rm eff}(t) = 0. 
\end{eqnarray}
This gives $\omega=-5/3$. Thus, the parameter $\omega\equiv p_{\rm eff}/\rho_{\rm eff}$ is not arbitrary and is uniquely determined to be $-5/3$. Using Eq.~\eqref{eq:divergence_K}, we obtain the relation:
\begin{eqnarray}
	\rho_{\rm eff}(t) = \rho_{\rm eff,0}\left(\frac{a}{a_0}\right)^2,
	\label{eq:rho_eff}
\end{eqnarray}
where $\rho_{\rm eff,0}$ represents the value at the present time. Thus, $\rho_{\rm eff}$ and $p_{\rm eff}$ can be regarded as the energy density and pressure for effective dark energy, respectively. Dividing Eq.~\eqref{eq:rho_eff} by $H_0^2$, we find that it represents the last term in Eq.~\eqref{eq:Hubble} with $\Omega_{\rm eff}\equiv \rho_{\rm eff,0}/H_0^2$. 
Once Eq.~\eqref{eq:rho_eff} is comprehended, we can employ Eqs.~\eqref{eq:Friedmann_a} and~\eqref{eq:Friedmann_b} as gravitational field equations.

What we have shown above is that Eq.~\eqref{eq:Harada_Friedmann3}, where dark energy is absent, is equivalent to Eqs.~\eqref{eq:Friedmann_a}--\eqref{eq:rho_eff}, where dark energy with $\omega=-5/3$ is included. This implies that we have two equivalent descriptions as follows.

The first description corresponds to the case in which we use only Eq.~\eqref{eq:Harada_Friedmann3} and do not utilize the Friedmann equations~\eqref{eq:Friedmann_a}--\eqref{eq:rho_eff}. In this case, the energy density $\rho$ consists of matter and radiation, $\rho=\rho_{\rm m}+\rho_{\rm r}$. The energy density for the effective component $\rho_{\rm eff}$ does not appear. Consequently, in this description, the concept of dark energy is absent and unnecessary. Using this first description, an expanding solution for the scale factor was discovered in a matter-dominated universe in Ref.~\cite{Harada:2023rqw}.

The second description is the case in which we use the Friedmann equations~\eqref{eq:Friedmann_a}--\eqref{eq:rho_eff}. In this case, the effective dark energy emerges. For practical applications, this second description is convenient because the Friedmann equations are widely recognized. However, the dark energy derived here should not be interpreted as an independent physical entity; it only emerges in the second description. In the first description, where Eq.~\eqref{eq:Harada_Friedmann3} is the sole equation in use, dark energy is absent. 

These two descriptions are physically equivalent. While the second description (where dark energy appears) is practically convenient, at least in principle, we can choose only the first description (where dark energy is absent). Therefore, dark energy derived here is an effective concept in the sense that it depends on the chosen descriptions. In contrast, for example, matter and radiation represent real physical entities, and they necessarily appear in both the first and second description.

For practical cosmological applications, we can choose the second description and begin with Eq.~\eqref{eq:Hubble} as follows. The additional term $\Omega_{\rm eff}(a/a_0)^2$ vanishes as $\Omega_{\rm eff}$ approaches zero. Consequently, when $\Omega_{\rm eff}$ gets very close to zero, Eq.~\eqref{eq:Hubble} simplifies to the standard Friedmann equation with no dark energy. However, if $\Omega_{\rm eff}$ assumes a small but nonzero value, it leads to deviations from standard cosmology. Furthermore, there is an even more intriguing possibility that $\Omega_{\rm eff}$ could assume a significant value with vanishing cosmological constant. 

In Sec.~\ref{sec:Hubble}, we will investigate the cosmological implications of both small and large values of $\Omega_{\rm eff}$.

\section{Cosmological implications\label{sec:Hubble}}

For cosmological applications, we begin with Eq.~\eqref{eq:Hubble}. 
It is convenient to express Eq.~\eqref{eq:Hubble} in terms of redshift:
\begin{eqnarray}
	\left(\frac{H(z)}{H_0}\right)^2
	=&& \Omega_{\rm m}(1+z)^3 + \Omega_{\rm r}(1+z)^4 + \Omega_k(1+z)^2\nonumber\\
	&&+  \Omega_\Lambda	
	+  \Omega_{\rm eff}(1+z)^{-2},
	\label{eq:Hubble3}
\end{eqnarray}
where $a(t)/a_0=1/(1+z)$. The five density parameters satisfy Eq.~\eqref{eq:five_parameters}.
It is also convenient to consider the quantity, $H(z)/(1+z)=\dot{a}(t)/a_0$, and its derivative. From Eq.~\eqref{eq:Hubble3}, we obtain
\begin{widetext}
\begin{eqnarray}
	\frac{H(z)}{1+z} =
	H_0 \sqrt{\Omega_{\rm m}(1+z)+ \Omega_{\rm r}(1+z)^2+\Omega_k + \Omega_\Lambda(1+z)^{-2} 
	+ \Omega_{\rm eff}(1+z)^{-4}},
	\label{eq:H/1+z}
\end{eqnarray}
and its derivative with respect to $z$ as
\begin{eqnarray}
	\frac{d}{dz}\left(\frac{H(z)}{1+z}\right)
	=H_0\frac{\frac{1}{2}\Omega_{\rm m}+\Omega_{\rm r}(1+z) - \Omega_\Lambda (1+z)^{-3}-2\Omega_{\rm eff}(1+z)^{-5}}{\sqrt{\Omega_{\rm m}(1+z)+ \Omega_{\rm r}(1+z)^2+\Omega_k + \Omega_\Lambda(1+z)^{-2} + \Omega_{\rm eff}(1+z)^{-4}}}.
	\label{eq:dH/1+z}
\end{eqnarray}
\end{widetext}

Here, the left-hand side of Eq.~\eqref{eq:dH/1+z} is calculated as
\begin{eqnarray}
	\frac{d}{dz}\left(\frac{H(z)}{1+z}\right) 
	= \frac{\dot{H}(1+z)/\dot{z}-H}{(1+z)^2},
	\label{eq:dH/1+z:2}
\end{eqnarray}
where dot denotes the time derivative. Substituting $\dot{z}=-(1+z)H$ into Eq.~\eqref{eq:dH/1+z:2} and using Eq.~\eqref{def:deceleration}, we obtain a useful formula:
\begin{eqnarray}
	\frac{d}{dz}\left(\frac{H(z)}{1+z}\right) = \frac{H(z)q(z)}{(1+z)^2}.
	\label{eq:slope}
\end{eqnarray}
Here, $q$ represents a deceleration parameter defined by Eq.~\eqref{def:deceleration}. From Eq.~\eqref{eq:slope}, we can see that the derivative, $d(H(z)/(1+z))/dz$, approaches $H_0 q_0$ as $z \rightarrow 0$. 

Equation~\eqref{eq:slope} indicates that when we plot $H(z)/(1+z)$ as a function of $z$, the slope is positive for decelerating expansion $(H>0, q>0)$, and negative for accelerating expansion $(H>0, q<0)$. The point where the transition from decelerating to accelerating expansion occurs corresponds to Eq.~\eqref{eq:slope} becoming zero. Although this can be easily deduced from the relation $H(z)/(1+z)=\dot{a}/a_0$, Eqs.~\eqref{eq:H/1+z}--\eqref{eq:slope} are convenient because they are expressed in terms of $\Omega$'s and $z$.

Substituting Eqs.~\eqref{eq:H/1+z} and~\eqref{eq:dH/1+z} into Eq.~\eqref{eq:slope}, we can express the deceleration parameter $q(z)$ in terms of $\Omega$'s as follows:
\begin{widetext}
\begin{eqnarray}
	q(z)
	=\frac{\frac{1}{2}\Omega_{\rm m}(1+z)^3+\Omega_{\rm r}(1+z)^4 - \Omega_\Lambda -2\Omega_{\rm eff}(1+z)^{-2}}{\Omega_{\rm m}(1+z)^3+ \Omega_{\rm r}(1+z)^4+\Omega_k (1+z)^2+ \Omega_\Lambda + \Omega_{\rm eff}(1+z)^{-2}}.
	\label{eq:deceleration}
\end{eqnarray}
\end{widetext}
The present deceleration parameter, denoted as $q_0$, can be obtained by substituting $z=0$ into Eq.~\eqref{eq:deceleration}:
\begin{eqnarray}
	q_0 	=\frac{1}{2}\Omega_{\rm m}+\Omega_{\rm r}-\Omega_\Lambda -2\Omega_{\rm eff}.
		\label{eq:q0_a}
\end{eqnarray}
In general relativity where $\Omega_{\rm eff}=0$, Eq.~\eqref{eq:q0_a} simplifies to $q_0=\Omega_{\rm m}/2+\Omega_{\rm r}-\Omega_{\rm \Lambda}$. In our theory, substituting $\Omega_{\rm eff}=1-\Omega_{\rm m}-\Omega_{\rm r}-\Omega_{\rm k}-\Omega_{\rm \Lambda}$ into Eq.~\eqref{eq:q0_a}, we obtain the relation:
\begin{eqnarray}
	q_0 	= \frac{5}{2}\Omega_{\rm m} + 3\Omega_{\rm r} + 2 \Omega_k + \Omega_\Lambda -2.
		\label{eq:q0}
\end{eqnarray}
This is consistent with Eq.~\eqref{eq:q0_Omega1}.

The transition redshift, denoted by $z_q$, is defined as the redshift at which the universe undergoes a transition from decelerating to accelerating expansion. It is determined by the condition $q(z=z_q)=0$, or equivalently, by the vanishing of Eq.~\eqref{eq:dH/1+z}. Substituting $z=z_q$ into Eq.~\eqref{eq:deceleration}, we obtain the condition that determines $z_q$ as
\begin{eqnarray}
	0&=&\frac{1}{2}\Omega_{\rm m}(1+z_q)^3 + \Omega_{\rm r}(1+z_q)^4 - \Omega_\Lambda \nonumber\\
	&&-2 \Omega_{\rm eff}(1+z_q)^{-2}.
	\label{eq:transition_redshift}
\end{eqnarray}

\begin{figure*}[htb]
\includegraphics[width=160mm]{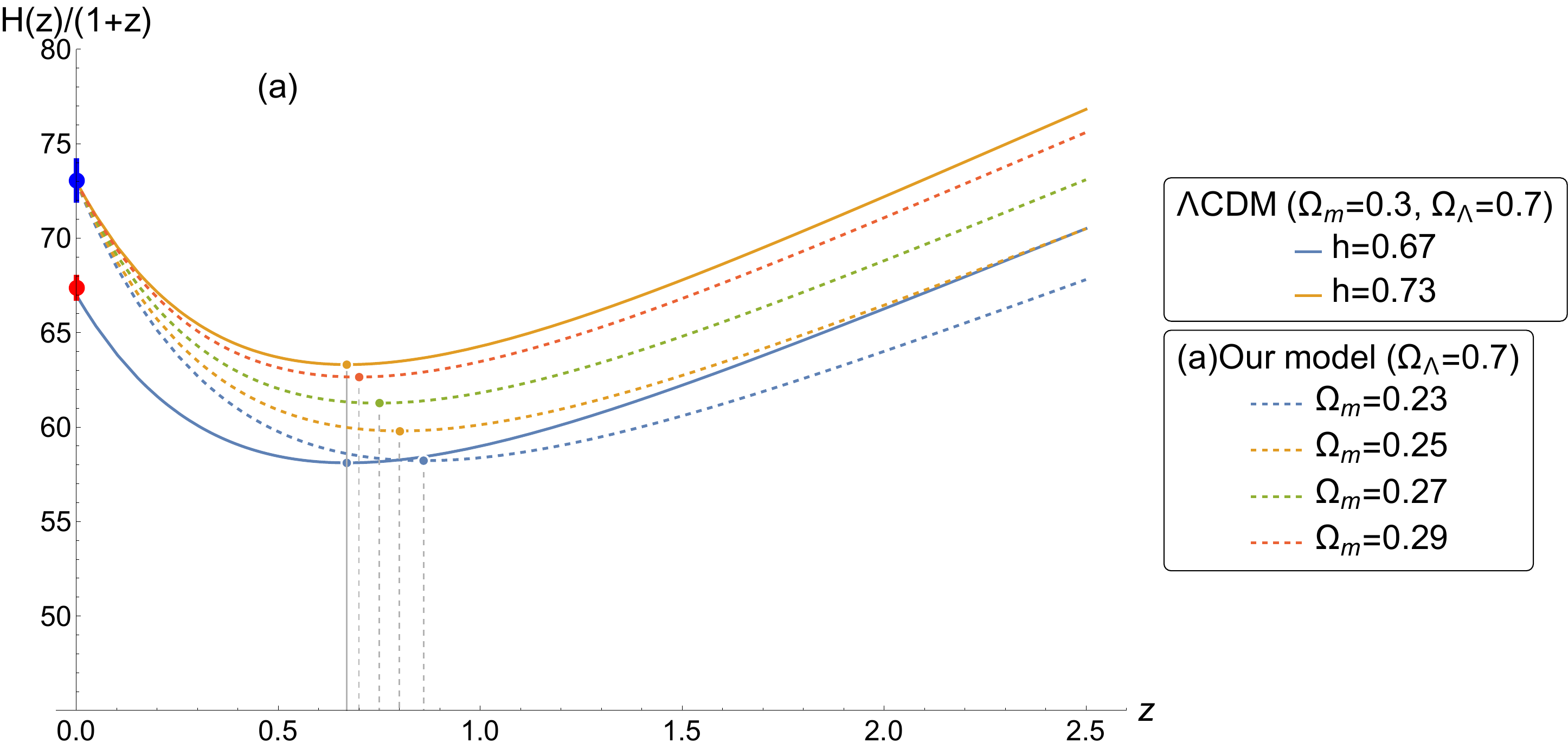}\\
\includegraphics[width=160mm]{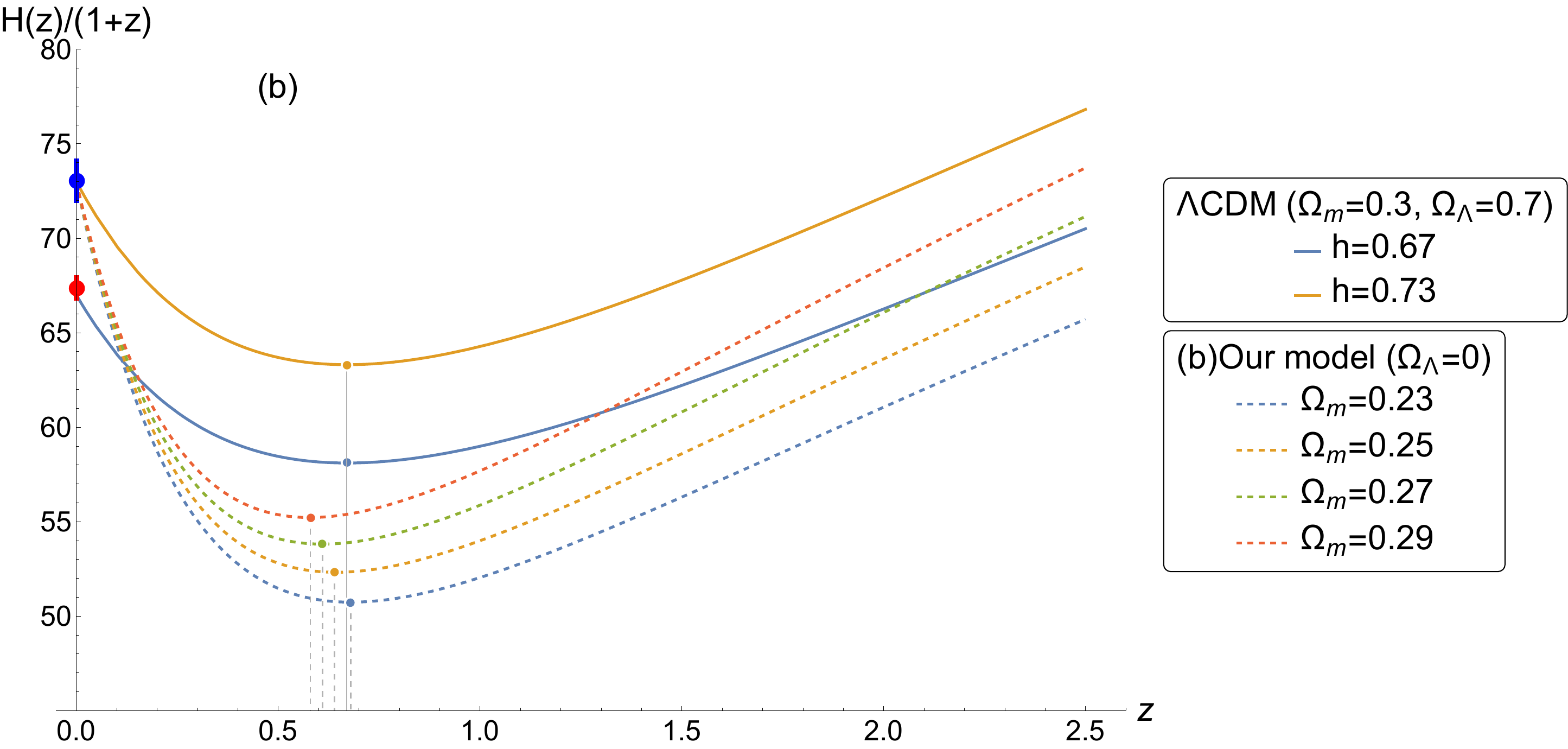}%
\caption{\label{fig:Hubble}
The $H(z)/(1+z)$ (in km s$^{-1}$ Mpc$^{-1}$) is plotted as a function of redshift. In all cases, $\Omega_k=\Omega_{\rm r}=0$ is assumed. In both panels (a) and (b), two solid curves represent the $\Lambda$CDM model with $\Omega_{\rm m}=0.3$, $\Omega_\Lambda=0.7$. The Hubble constant, $H_0=100 h$ km s$^{-1}$ Mpc$^{-1}$,  is assumed to be $h=0.67$ for the lower solid curve (blue), and $h=0.73$ for the upper solid curve (orange), respectively. The four dashed curves in both panels represent the cases including the effective dark energy defined by $\Omega_{\rm eff}\equiv 1 - \Omega_{\rm m}-\Omega_\Lambda$: panel (a) represents the case with $\Omega_\Lambda=0.7$, and panel (b) represents the case with $\Omega_\Lambda=0$. The blue point with bar at $z=0$ represents $H_0=73.0 \pm 1.0$ km s$^{-1}$ Mpc$^{-1}$ obtained from the local distance measurements~\cite{Riess:2021jrx}, while the red point with bar at $z=0$ represents $H_0=67.4 \pm 0.5$ km s$^{-1}$ Mpc$^{-1}$ obtained from the Planck CMB data~\cite{Planck:2018vyg}. The slope of the curves at $z=0$ represents $H_0 q_0$. The points on the curves at $z\sim 0.6$ represent the cosmological transition point from decelerating to accelerating expansion. Using Eq.~\eqref{eq:q0} and~\eqref{eq:transition_redshift}, we obtain $q_0=-0.55$ and $z_q\simeq 0.67$ for the two solid curves in both panels. In panel (a), for the four dashed curves with $\Omega_{\rm m}=(0.23, 0.25, 0.27, 0.29)$, we obtain $q_0=(-0.725, -0.675, -0.625, -0.575)$ and $z_q=(0.86, 0.80, 0.75, 0.70)$. In panel (b), for the four dashed curves with $\Omega_{\rm m}=(0.23, 0.25, 0.27, 0.29)$, we obtain $q_0=(-1.425, -1.375, -1.325, -1.275)$ and $z_q=(0.68, 0.64, 0.61, 0.58)$.}
\end{figure*}

Figure~\ref{fig:Hubble} illustrates $H(z)/(1+z)$ as a function of redshift. We explore two distinct cosmological models: panel (a) in Fig.~\ref{fig:Hubble} represents the model with $\Omega_\Lambda=0.7$, while panel (b) in Fig.~\ref{fig:Hubble} illustrates the model with $\Omega_\Lambda=0$.

The model presented in panel (a) of Fig.~\ref{fig:Hubble} assumes a small value for $\Omega_{\rm eff}$ and a large value for $\Omega_\Lambda=0.7$. Consequently, this model exhibits relatively minor deviations from the standard $\Lambda$CDM model, and the current accelerating expansion is attributed to the cosmological constant. Panel (a) in Fig.~\ref{fig:Hubble} demonstrates that this model has the potential to address the Hubble tension problem. In this model, as described in Fig.~\ref{fig:Hubble}'s caption, the transition redshift $z_q$ has a larger value than the $\Lambda$CDM value of $z_q\simeq0.68$. Additionally, the current deceleration parameter $|q_0|$ exceeds the $\Lambda$CDM value of $|q_0|\simeq0.55$. 

This model appears to address the Hubble tension. However, it does not resolve the coincidence problem associated with the cosmological constant. While the energy density of matter follows $\rho_{\rm m} \propto a^{-3}$, $\Lambda$ remains constant. These two components, matter and $\Lambda$, represent distinct physical entities that are independent of each other. Nevertheless, $\Lambda$ must be adjusted to be $\Omega_\Lambda \sim \Omega_{\rm m} \sim {\cal O}(1)$ as an initial condition. The necessity for such tuning is theoretically unsatisfactory, leading us to consider the second model.

The model presented in panel (b) of Fig.~\ref{fig:Hubble} assumes $\Omega_\Lambda=0$. In this model, only matter exists as an independent physical entity. The effective component $\Omega_{\rm eff}$ is given by $\Omega_{\rm eff}=1-\Omega_{\rm m}$ in the present case, and hence it is not independent of $\Omega_{\rm m}$. Indeed, if we choose the first description mentioned in Sec.~\ref{sec:omega} in which only Eq.~\eqref{eq:Harada_Friedmann3} is used, $\Omega_{\rm eff}$ does not appear. Only $\Omega_{\rm m}$ is a unique independent component. Consequently, the coincidence problem is resolved in this model; there is no need for any tuning. Theoretically, this presents an advantage compared to model shown in panel (a). 

Panel (b) in Fig.~\ref{fig:Hubble} indicates that this model also has the potential to address the Hubble tension problem. While the transition redshift $z_q$ has a similar value to that of $\Lambda$CDM, the value of $H(z)/(1+z)$ might appear small at $z\sim 0.6$, resulting in large values of $|q_0|$. At present, however, there still remains a large uncertainty in the observational values. 

\begin{figure*}[htb]
\includegraphics[width=170mm]{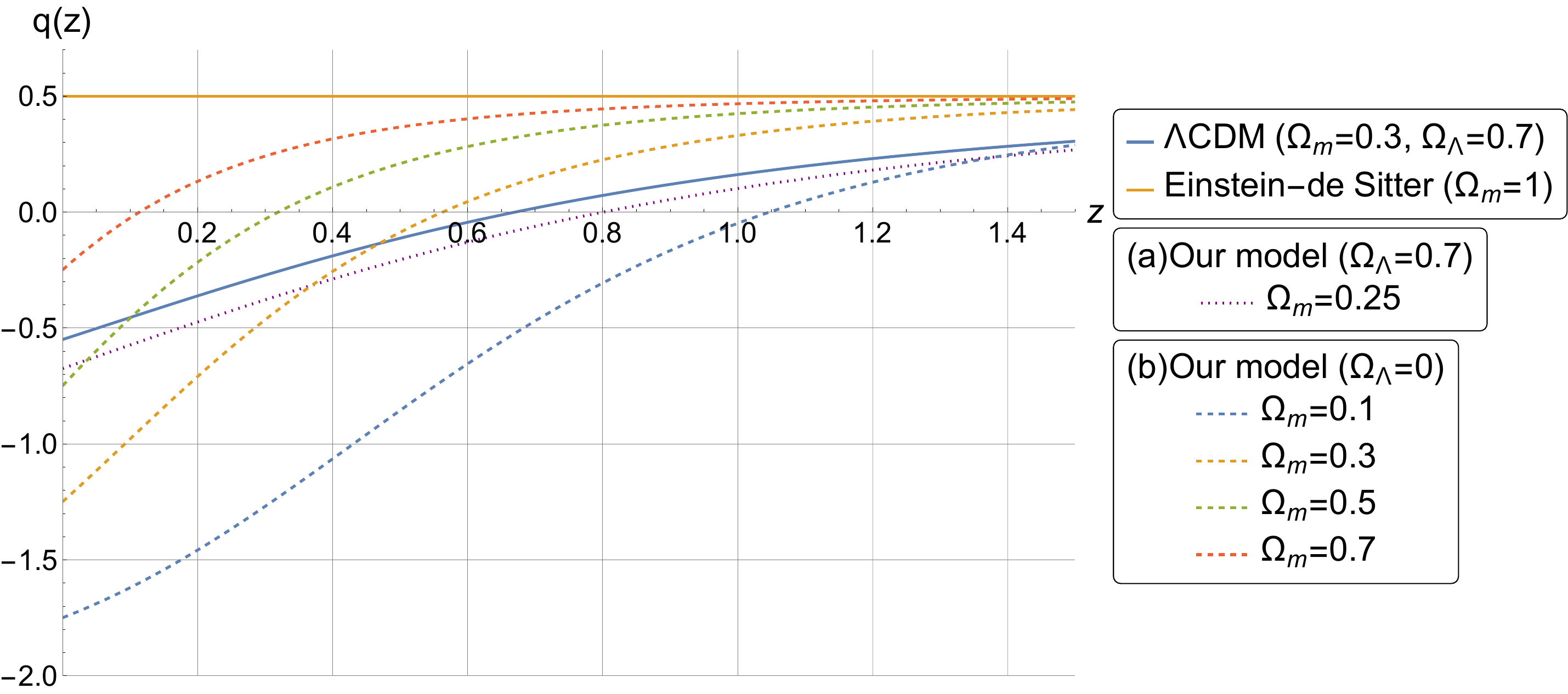}%
\caption{\label{fig:deceleration} The deceleration parameter $q(z)$ is shown as a function of redshift. In all cases, $\Omega_k=\Omega_{\rm r}=0$ is assumed. The dotted curve and the four dashed curves represent our models including effective dark energy defined by $\Omega_{\rm eff}\equiv 1-\Omega_{\rm m}-\Omega_\Lambda$: the dotted curve represents the model (a) with $\Omega_\Lambda=0.7$, and the four dashed curves represent the model (b) with $\Omega_\Lambda=0$. The  solid curve (blue) represents the $\Lambda$CDM with $\Omega_{\rm m}=0.3$ and $\Omega_\Lambda=0.7$. The solid line (orange) represents the Einstein--de Sitter model. The points where each curve crosses the horizontal $z$ axis represent the transition redshift $z_q$ from decelerating to accelerating expansion. The points at $z=0$ represent the present-time value of the deceleration parameter $q_0$.}
\end{figure*}

Figure~\ref{fig:deceleration} illustrates the deceleration parameter $q(z)$ as a function of redshift. This figure demonstrates that the behavior of $q(z)$ for model (b) with $\Omega_\Lambda=0$ differs from that of $\Lambda$CDM. Once the values $q_0$ and $z_q$ will be precisely determined by future observations, it could potentially help in distinguishing  between model (b) and $\Lambda$CDM. 

Finally, using Eq.~\eqref{eq:Hubble3} and following Ref.~\cite{Weinberg:2008zzc}, we can derive the expression for the luminosity distance $d_{\rm L}(z)$ of an observed source:
\begin{widetext}
\begin{eqnarray}
	d_{\rm L}(z) =\frac{1+z}{H_0\sqrt{\Omega_k}}\sinh \left[ \sqrt{\Omega_k}\int_{\frac{1}{1+z}}^1  \frac{dx}{x^2 \sqrt{\Omega_{\rm m}x^{-3} + \Omega_{\rm r}x^{-4} + \Omega_k x^{-2} +\Omega_\Lambda+ \Omega_{\rm eff}x^2}}\right],
\end{eqnarray}
which can be used for any $\Omega_k$. For $\Omega_k=\Omega_{\rm r}=0$, the expression is given by
\begin{eqnarray}
	d_{\rm L}(z) = \frac{1+z}{H_0}\int_{\frac{1}{1+z}}^1
	\frac{dx}{x^2\sqrt{\Omega_{\rm m}x^{-3} + \Omega_\Lambda + (1-\Omega_{\rm m}-\Omega_\Lambda)x^2}},\qquad
	\label{eq:dL2}
\end{eqnarray}
\end{widetext}
where the $\Omega$'s satisfy the relation, $5\Omega_{\rm m}/2+\Omega_\Lambda=2+q_0$. 

\begin{figure*}[t]
\includegraphics[width=170mm]{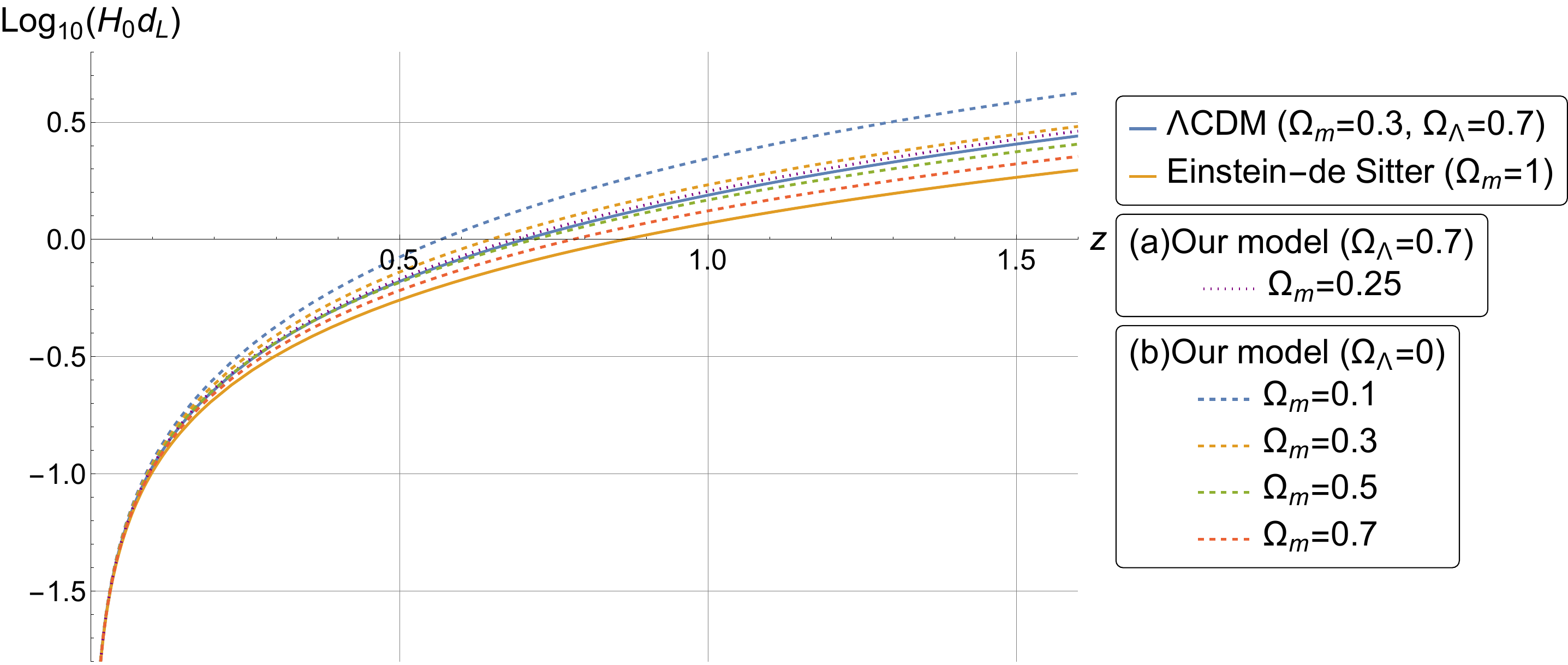}\\
\caption{\label{fig:luminosity_disntace} The Hubble constant-free luminosity distance, $\log_{10} (H_0 d_{\rm L})$, is plotted as a function of redshift. In all cases, $\Omega_k=\Omega_{\rm r}=0$ is assumed. All curves correspond to those shown in Fig.~\ref{fig:deceleration}. The dotted curve and the dashed curves for $\Omega_{\rm m}\leq 0.5$ can be distinguished from the Einstein--de Sitter model.}
\end{figure*}

Figure~\ref{fig:luminosity_disntace} illustrates the Hubble constant-free luminosity distance, $\log_{10} (H_0 d_{\rm L}(z))$, as a function of redshift. This figure demonstrates that even if $\Omega_\Lambda=0$, our model for $\Omega_{\rm m} \leq 0.5$ can be distinguished from the decelerating Einstein--de Sitter model. This suggests that a model with $\Lambda=0$ can be a viable cosmological model to describe the present-day universe.

\section{Summary and conclusions\label{sec:conclusion}}
We have demonstrated that the recently derived evolution equation for the scale factor, Eq.~\eqref{eq:Harada_Friedmann3}, with no dark energy, is equivalent to the standard Friedmann equation which includes a specific type of dark energy. Consequently, there is no need to assume the existence of dark energy as a separate physical entity. The effective dark energy derived in this work is characterized by an equation of state parameter, $\omega=-5/3$, which is solely determined by the gravitational field equation~\eqref{eq:Harada_Friedmann3}. 

As depicted in Fig.~\ref{fig:Hubble}, our findings demonstrate that when the effective dark energy ($\Omega_{\rm eff}\equiv 1-\Omega_{\rm m}-\Omega_\Lambda$) is present in a moderate amount, typically around 5\% of the total energy density, it holds the potential to resolve the Hubble tension issue. Furthermore, even if the effective dark energy is dominant, typically around $70\%$ of the total energy density with $\Lambda=0$, it still holds the potential to address the Hubble tension. The latter case resolves the coincidence problem related to the cosmological constant, because only $\Omega_{\rm m}$ is a unique independent component. This is theoretically an advantage than the models with nonzero $\Lambda$. As shown in Fig.~\ref{fig:deceleration}, the effective dark energy influences the deceleration parameter $q(z)$ and the transition redshift $z_q$. Precisely determining these parameters could help in distinguishing whether the energy density of effective dark energy is zero or nonzero. 

\noindent {\it Note added:} While completing this paper, the author received a paper by C.~A.~Mantica and L.~G.~Molinari~\cite{Mantica:2023} which also reports that an equation of state parameter for the additional component is determined to be $-5/3$. 

\begin{center}
{\bf ACKNOWLEDGMENTS}\\
\end{center}

This work was supported by JSPS KAKENHI Grant No. JP22K03599.


\bibliography{references}
\end{document}